\documentclass[11pt,a4paper,english,superscriptaddress,aps]{revtex4}
\usepackage{graphicx}
\makeatletter
\usepackage{babel}
\usepackage[latin1]{inputenc}
\usepackage{hyperref}
\renewcommand{\a}{\alpha}
\renewcommand{\b}{\beta}

\def\m{\mu}

\def\r{\rho}

\def\t{\tau}

\def\x{\xi}

\def\L{\Lambda}

\newcommand{\bb}{\bibitem}
\def\pls{\partial\!\!\!/}
\def\bb{\bibitem}

\def\ps{p\!\!\!/}
\def\bs{b\!\!\!/}

\def\n{\nu}
\def\m{\mu}
\def\n{\nu}
\def\bb{\bibitem}

\newcommand{\be}{\begin{equation}}
\newcommand{\ee}{\end{equation}}
\newcommand{\bea}{\begin{eqnarray}}
\newcommand{\eea}{\end{eqnarray}}

\newcommand{\pa}{\partial}

\begin{document}

\immediate\write16{<WARNING: FEYNMAN macros work only with emTeX-dvivers
                    (dviscr.exe, dvihplj.exe, dvidot.exe, etc.) >}
\newdimen\Lengthunit
\newcount\Nhalfperiods
\Lengthunit = 1.5cm
\Nhalfperiods = 9
\catcode`\*=11
\newdimen\L*   \newdimen\d*   \newdimen\d**
\newdimen\dm*  \newdimen\dd*  \newdimen\dt*
\newdimen\a*   \newdimen\b*   \newdimen\c*
\newdimen\a**  \newdimen\b**
\newdimen\xL*  \newdimen\yL*
\newcount\k*   \newcount\l*   \newcount\m*
\newcount\n*   \newcount\dn*  \newcount\r*
\newcount\N*   \newcount\*one \newcount\*two  \*one=1 \*two=2
\newcount\*ths \*ths=1000
\def\GRAPH(hsize=#1)#2{\hbox to #1\Lengthunit{#2\hss}}
\def\Linewidth#1{\special{em:linewidth #1}}
\Linewidth{.4pt}
\def\sm*{\special{em:moveto}}
\def\sl*{\special{em:lineto}}
\newbox\spm*   \newbox\spl*
\setbox\spm*\hbox{\sm*}
\setbox\spl*\hbox{\sl*}
\def\mov#1(#2,#3)#4{\rlap{\L*=#1\Lengthunit\kern#2\L*\raise#3\L*\hbox{#4}}}
\def\smov#1(#2,#3)#4{\rlap{\L*=#1\Lengthunit
\xL*=\xscale\L*\yL*=\yscale\L*\kern#2\xL*\raise#3\yL*\hbox{#4}}}
\def\mov*(#1,#2)#3{\rlap{\kern#1\raise#2\hbox{#3}}}
\def\lin#1(#2,#3){\rlap{\sm*\mov#1(#2,#3){\sl*}}}
\def\arr*(#1,#2,#3){\mov*(#1\dd*,#1\dt*){%
\sm*\mov*(#2\dd*,#2\dt*){\mov*(#3\dt*,-#3\dd*){\sl*}}%
\sm*\mov*(#2\dd*,#2\dt*){\mov*(-#3\dt*,#3\dd*){\sl*}}}}
\def\arrow#1(#2,#3){\rlap{\lin#1(#2,#3)\mov#1(#2,#3){%
\d**=-.012\Lengthunit\dd*=#2\d**\dt*=#3\d**%
\arr*(1,10,4)\arr*(3,8,4)\arr*(4.8,4.2,3)}}}
\def\arrlin#1(#2,#3){\rlap{\L*=#1\Lengthunit\L*=.5\L*%
\lin#1(#2,#3)\mov*(#2\L*,#3\L*){\arrow.1(#2,#3)}}}
\def\dasharrow#1(#2,#3){\rlap{%
{\Lengthunit=0.9\Lengthunit\dashlin#1(#2,#3)\mov#1(#2,#3){\sm*}}%
\mov#1(#2,#3){\sl*\d**=-.012\Lengthunit\dd*=#2\d**\dt*=#3\d**%
\arr*(1,10,4)\arr*(3,8,4)\arr*(4.8,4.2,3)}}}
\def\clap#1{\hbox to 0pt{\hss #1\hss}}
\def\ind(#1,#2)#3{\rlap{%
\d*=.1\Lengthunit\kern#1\d*\raise#2\d*\hbox{\lower2pt\clap{$#3$}}}}
\def\sh*(#1,#2)#3{\rlap{%
\dm*=\the\n*\d**\xL*=\xscale\dm*\yL*=\yscale\dm*
\kern#1\xL*\raise#2\yL*\hbox{#3}}}
\def\calcnum*#1(#2,#3){\a*=1000sp\b*=1000sp\a*=#2\a*\b*=#3\b*%
\ifdim\a*<0pt\a*-\a*\fi\ifdim\b*<0pt\b*-\b*\fi%
\ifdim\a*>\b*\c*=.96\a*\advance\c*.4\b*%
\else\c*=.96\b*\advance\c*.4\a*\fi%
\k*\a*\multiply\k*\k*\l*\b*\multiply\l*\l*%
\m*\k*\advance\m*\l*\n*\c*\r*\n*\multiply\n*\n*%
\dn*\m*\advance\dn*-\n*\divide\dn*2\divide\dn*\r*%
\advance\r*\dn*%
\c*=\the\Nhalfperiods5sp\c*=#1\c*\ifdim\c*<0pt\c*-\c*\fi%
\multiply\c*\r*\N*\c*\divide\N*10000}
\def\dashlin#1(#2,#3){\rlap{\calcnum*#1(#2,#3)%
\d**=#1\Lengthunit\ifdim\d**<0pt\d**-\d**\fi%
\divide\N*2\multiply\N*2\advance\N*1%
\divide\d**\N*\sm*\n*\*one\sh*(#2,#3){\sl*}%
\loop\advance\n*\*one\sh*(#2,#3){\sm*}\advance\n*\*one\sh*(#2,#3){\sl*}%
\ifnum\n*<\N*\repeat}}
\def\dashdotlin#1(#2,#3){\rlap{\calcnum*#1(#2,#3)%
\d**=#1\Lengthunit\ifdim\d**<0pt\d**-\d**\fi%
\divide\N*2\multiply\N*2\advance\N*1\multiply\N*2%
\divide\d**\N*\sm*\n*\*two\sh*(#2,#3){\sl*}\loop%
\advance\n*\*one\sh*(#2,#3){\kern-1.48pt\lower.5pt\hbox{\rm.}}%
\advance\n*\*one\sh*(#2,#3){\sm*}%
\advance\n*\*two\sh*(#2,#3){\sl*}\ifnum\n*<\N*\repeat}}
\def\shl*(#1,#2)#3{\kern#1#3\lower#2#3\hbox{\unhcopy\spl*}}
\def\trianglin#1(#2,#3){\rlap{\toks0={#2}\toks1={#3}\calcnum*#1(#2,#3)%
\dd*=.57\Lengthunit\dd*=#1\dd*\divide\dd*\N*%
\d**=#1\Lengthunit\ifdim\d**<0pt\d**-\d**\fi%
\multiply\N*2\divide\d**\N*\advance\N*-1\sm*\n*\*one\loop%
\shl**{\dd*}\dd*-\dd*\advance\n*2%
\ifnum\n*<\N*\repeat\n*\N*\advance\n*1\shl**{0pt}}}
\def\wavelin#1(#2,#3){\rlap{\toks0={#2}\toks1={#3}\calcnum*#1(#2,#3)%
\dd*=.23\Lengthunit\dd*=#1\dd*\divide\dd*\N*%
\d**=#1\Lengthunit\ifdim\d**<0pt\d**-\d**\fi%
\multiply\N*4\divide\d**\N*\sm*\n*\*one\loop%
\shl**{\dd*}\dt*=1.3\dd*\advance\n*1%
\shl**{\dt*}\advance\n*\*one%
\shl**{\dd*}\advance\n*\*two%
\dd*-\dd*\ifnum\n*<\N*\repeat\n*\N*\shl**{0pt}}}
\def\w*lin(#1,#2){\rlap{\toks0={#1}\toks1={#2}\d**=\Lengthunit\dd*=-.12\d**%
\N*8\divide\d**\N*\sm*\n*\*one\loop%
\shl**{\dd*}\dt*=1.3\dd*\advance\n*\*one%
\shl**{\dt*}\advance\n*\*one%
\shl**{\dd*}\advance\n*\*one%
\shl**{0pt}\dd*-\dd*\advance\n*1\ifnum\n*<\N*\repeat}}
\def\l*arc(#1,#2)[#3][#4]{\rlap{\toks0={#1}\toks1={#2}\d**=\Lengthunit%
\dd*=#3.037\d**\dd*=#4\dd*\dt*=#3.049\d**\dt*=#4\dt*\ifdim\d**>16mm%
\d**=.25\d**\n*\*one\shl**{-\dd*}\n*\*two\shl**{-\dt*}\n*3\relax%
\shl**{-\dd*}\n*4\relax\shl**{0pt}\else\ifdim\d**>5mm%
\d**=.5\d**\n*\*one\shl**{-\dt*}\n*\*two\shl**{0pt}%
\else\n*\*one\shl**{0pt}\fi\fi}}
\def\d*arc(#1,#2)[#3][#4]{\rlap{\toks0={#1}\toks1={#2}\d**=\Lengthunit%
\dd*=#3.037\d**\dd*=#4\dd*\d**=.25\d**\sm*\n*\*one\shl**{-\dd*}%
\n*3\relax\sh*(#1,#2){\xL*=\xscale\dd*\yL*=\yscale\dd*
\kern#2\xL*\lower#1\yL*\hbox{\sm*}}%
\n*4\relax\shl**{0pt}}}
\def\arc#1[#2][#3]{\rlap{\Lengthunit=#1\Lengthunit%
\sm*\l*arc(#2.1914,#3.0381)[#2][#3]%
\smov(#2.1914,#3.0381){\l*arc(#2.1622,#3.1084)[#2][#3]}%
\smov(#2.3536,#3.1465){\l*arc(#2.1084,#3.1622)[#2][#3]}%
\smov(#2.4619,#3.3086){\l*arc(#2.0381,#3.1914)[#2][#3]}}}
\def\dasharc#1[#2][#3]{\rlap{\Lengthunit=#1\Lengthunit%
\d*arc(#2.1914,#3.0381)[#2][#3]%
\smov(#2.1914,#3.0381){\d*arc(#2.1622,#3.1084)[#2][#3]}%
\smov(#2.3536,#3.1465){\d*arc(#2.1084,#3.1622)[#2][#3]}%
\smov(#2.4619,#3.3086){\d*arc(#2.0381,#3.1914)[#2][#3]}}}
\def\wavearc#1[#2][#3]{\rlap{\Lengthunit=#1\Lengthunit%
\w*lin(#2.1914,#3.0381)%
\smov(#2.1914,#3.0381){\w*lin(#2.1622,#3.1084)}%
\smov(#2.3536,#3.1465){\w*lin(#2.1084,#3.1622)}%
\smov(#2.4619,#3.3086){\w*lin(#2.0381,#3.1914)}}}
\def\shl**#1{\c*=\the\n*\d**\d*=#1%
\a*=\the\toks0\c*\b*=\the\toks1\d*\advance\a*-\b*%
\b*=\the\toks1\c*\d*=\the\toks0\d*\advance\b*\d*%
\a*=\xscale\a*\b*=\yscale\b*%
\raise\b*\rlap{\kern\a*\unhcopy\spl*}}
\def\wlin*#1(#2,#3)[#4]{\rlap{\toks0={#2}\toks1={#3}%
\c*=#1\l*\c*\c*=.01\Lengthunit\m*\c*\divide\l*\m*%
\c*=\the\Nhalfperiods5sp\multiply\c*\l*\N*\c*\divide\N*\*ths%
\divide\N*2\multiply\N*2\advance\N*1%
\dd*=.002\Lengthunit\dd*=#4\dd*\multiply\dd*\l*\divide\dd*\N*%
\d**=#1\multiply\N*4\divide\d**\N*\sm*\n*\*one\loop%
\shl**{\dd*}\dt*=1.3\dd*\advance\n*\*one%
\shl**{\dt*}\advance\n*\*one%
\shl**{\dd*}\advance\n*\*two%
\dd*-\dd*\ifnum\n*<\N*\repeat\n*\N*\shl**{0pt}}}
\def\wavebox#1{\setbox0\hbox{#1}%
\a*=\wd0\advance\a*14pt\b*=\ht0\advance\b*\dp0\advance\b*14pt%
\hbox{\kern9pt%
\mov*(0pt,\ht0){\mov*(-7pt,7pt){\wlin*\a*(1,0)[+]\wlin*\b*(0,-1)[-]}}%
\mov*(\wd0,-\dp0){\mov*(7pt,-7pt){\wlin*\a*(-1,0)[+]\wlin*\b*(0,1)[-]}}%
\box0\kern9pt}}
\def\rectangle#1(#2,#3){%
\lin#1(#2,0)\lin#1(0,#3)\mov#1(0,#3){\lin#1(#2,0)}\mov#1(#2,0){\lin#1(0,#3)}}
\def\dashrectangle#1(#2,#3){\dashlin#1(#2,0)\dashlin#1(0,#3)%
\mov#1(0,#3){\dashlin#1(#2,0)}\mov#1(#2,0){\dashlin#1(0,#3)}}
\def\waverectangle#1(#2,#3){\L*=#1\Lengthunit\a*=#2\L*\b*=#3\L*%
\ifdim\a*<0pt\a*-\a*\def\x*{-1}\else\def\x*{1}\fi%
\ifdim\b*<0pt\b*-\b*\def\y*{-1}\else\def\y*{1}\fi%
\wlin*\a*(\x*,0)[-]\wlin*\b*(0,\y*)[+]%
\mov#1(0,#3){\wlin*\a*(\x*,0)[+]}\mov#1(#2,0){\wlin*\b*(0,\y*)[-]}}
\def\calcparab*{%
\ifnum\n*>\m*\k*\N*\advance\k*-\n*\else\k*\n*\fi%
\a*=\the\k* sp\a*=10\a*\b*\dm*\advance\b*-\a*\k*\b*%
\a*=\the\*ths\b*\divide\a*\l*\multiply\a*\k*%
\divide\a*\l*\k*\*ths\r*\a*\advance\k*-\r*%
\dt*=\the\k*\L*}
\def\arcto#1(#2,#3)[#4]{\rlap{\toks0={#2}\toks1={#3}\calcnum*#1(#2,#3)%
\dm*=135sp\dm*=#1\dm*\d**=#1\Lengthunit\ifdim\dm*<0pt\dm*-\dm*\fi%
\multiply\dm*\r*\a*=.3\dm*\a*=#4\a*\ifdim\a*<0pt\a*-\a*\fi%
\advance\dm*\a*\N*\dm*\divide\N*10000%
\divide\N*2\multiply\N*2\advance\N*1%
\L*=-.25\d**\L*=#4\L*\divide\d**\N*\divide\L*\*ths%
\m*\N*\divide\m*2\dm*=\the\m*5sp\l*\dm*%
\sm*\n*\*one\loop\calcparab*\shl**{-\dt*}%
\advance\n*1\ifnum\n*<\N*\repeat}}
\def\arrarcto#1(#2,#3)[#4]{\L*=#1\Lengthunit\L*=.54\L*%
\arcto#1(#2,#3)[#4]\mov*(#2\L*,#3\L*){\d*=.457\L*\d*=#4\d*\d**-\d*%
\mov*(#3\d**,#2\d*){\arrow.02(#2,#3)}}}
\def\dasharcto#1(#2,#3)[#4]{\rlap{\toks0={#2}\toks1={#3}\calcnum*#1(#2,#3)%
\dm*=\the\N*5sp\a*=.3\dm*\a*=#4\a*\ifdim\a*<0pt\a*-\a*\fi%
\advance\dm*\a*\N*\dm*%
\divide\N*20\multiply\N*2\advance\N*1\d**=#1\Lengthunit%
\L*=-.25\d**\L*=#4\L*\divide\d**\N*\divide\L*\*ths%
\m*\N*\divide\m*2\dm*=\the\m*5sp\l*\dm*%
\sm*\n*\*one\loop%
\calcparab*\shl**{-\dt*}\advance\n*1%
\ifnum\n*>\N*\else\calcparab*%
\sh*(#2,#3){\kern#3\dt*\lower#2\dt*\hbox{\sm*}}\fi%
\advance\n*1\ifnum\n*<\N*\repeat}}
\def\*shl*#1{%
\c*=\the\n*\d**\advance\c*#1\a**\d*\dt*\advance\d*#1\b**%
\a*=\the\toks0\c*\b*=\the\toks1\d*\advance\a*-\b*%
\b*=\the\toks1\c*\d*=\the\toks0\d*\advance\b*\d*%
\raise\b*\rlap{\kern\a*\unhcopy\spl*}}
\def\calcnormal*#1{%
\b**=10000sp\a**\b**\k*\n*\advance\k*-\m*%
\multiply\a**\k*\divide\a**\m*\a**=#1\a**\ifdim\a**<0pt\a**-\a**\fi%
\ifdim\a**>\b**\d*=.96\a**\advance\d*.4\b**%
\else\d*=.96\b**\advance\d*.4\a**\fi%
\d*=.01\d*\r*\d*\divide\a**\r*\divide\b**\r*%
\ifnum\k*<0\a**-\a**\fi\d*=#1\d*\ifdim\d*<0pt\b**-\b**\fi%
\k*\a**\a**=\the\k*\dd*\k*\b**\b**=\the\k*\dd*}
\def\wavearcto#1(#2,#3)[#4]{\rlap{\toks0={#2}\toks1={#3}\calcnum*#1(#2,#3)%
\c*=\the\N*5sp\a*=.4\c*\a*=#4\a*\ifdim\a*<0pt\a*-\a*\fi%
\advance\c*\a*\N*\c*\divide\N*20\multiply\N*2\advance\N*-1\multiply\N*4%
\d**=#1\Lengthunit\dd*=.012\d**\ifdim\d**<0pt\d**-\d**\fi\L*=.25\d**%
\divide\d**\N*\divide\dd*\N*\L*=#4\L*\divide\L*\*ths%
\m*\N*\divide\m*2\dm*=\the\m*0sp\l*\dm*%
\sm*\n*\*one\loop\calcnormal*{#4}\calcparab*%
\*shl*{1}\advance\n*\*one\calcparab*%
\*shl*{1.3}\advance\n*\*one\calcparab*%
\*shl*{1}\advance\n*2%
\dd*-\dd*\ifnum\n*<\N*\repeat\n*\N*\shl**{0pt}}}
\def\triangarcto#1(#2,#3)[#4]{\rlap{\toks0={#2}\toks1={#3}\calcnum*#1(#2,#3)%
\c*=\the\N*5sp\a*=.4\c*\a*=#4\a*\ifdim\a*<0pt\a*-\a*\fi%
\advance\c*\a*\N*\c*\divide\N*20\multiply\N*2\advance\N*-1\multiply\N*2%
\d**=#1\Lengthunit\dd*=.012\d**\ifdim\d**<0pt\d**-\d**\fi\L*=.25\d**%
\divide\d**\N*\divide\dd*\N*\L*=#4\L*\divide\L*\*ths%
\m*\N*\divide\m*2\dm*=\the\m*0sp\l*\dm*%
\sm*\n*\*one\loop\calcnormal*{#4}\calcparab*%
\*shl*{1}\advance\n*2%
\dd*-\dd*\ifnum\n*<\N*\repeat\n*\N*\shl**{0pt}}}
\def\hr*#1{\clap{\xL*=\xscale\Lengthunit\vrule width#1\xL* height.1pt}}
\def\shade#1[#2]{\rlap{\Lengthunit=#1\Lengthunit%
\smov(0,#2.05){\hr*{.994}}\smov(0,#2.1){\hr*{.980}}%
\smov(0,#2.15){\hr*{.953}}\smov(0,#2.2){\hr*{.916}}%
\smov(0,#2.25){\hr*{.867}}\smov(0,#2.3){\hr*{.798}}%
\smov(0,#2.35){\hr*{.715}}\smov(0,#2.4){\hr*{.603}}%
\smov(0,#2.45){\hr*{.435}}}}
\def\dshade#1[#2]{\rlap{%
\Lengthunit=#1\Lengthunit\if#2-\def\t*{+}\else\def\t*{-}\fi%
\smov(0,\t*.025){%
\smov(0,#2.05){\hr*{.995}}\smov(0,#2.1){\hr*{.988}}%
\smov(0,#2.15){\hr*{.969}}\smov(0,#2.2){\hr*{.937}}%
\smov(0,#2.25){\hr*{.893}}\smov(0,#2.3){\hr*{.836}}%
\smov(0,#2.35){\hr*{.760}}\smov(0,#2.4){\hr*{.662}}%
\smov(0,#2.45){\hr*{.531}}\smov(0,#2.5){\hr*{.320}}}}}
\def\vdot{\rlap{\kern-1.9pt\lower1.8pt\hbox{$\scriptstyle\bullet$}}}
\def\vtimes{\rlap{\kern-3pt\lower1.8pt\hbox{$\scriptstyle\times$}}}
\def\vDot{\rlap{\kern-2.3pt\lower2.7pt\hbox{$\bullet$}}}
\def\vTimes{\rlap{\kern-3.6pt\lower2.4pt\hbox{$\times$}}}
\catcode`\*=12
\newcount\CatcodeOfAtSign
\CatcodeOfAtSign=\the\catcode`\@
\catcode`\@=11
\newcount\n@ast
\def\n@ast@#1{\n@ast0\relax\get@ast@#1\end}
\def\get@ast@#1{\ifx#1\end\let\next\relax\else%
\ifx#1*\advance\n@ast1\fi\let\next\get@ast@\fi\next}
\newif\if@up \newif\if@dwn
\def\up@down@#1{\@upfalse\@dwnfalse%
\if#1u\@uptrue\fi\if#1U\@uptrue\fi\if#1+\@uptrue\fi%
\if#1d\@dwntrue\fi\if#1D\@dwntrue\fi\if#1-\@dwntrue\fi}
\def\halfcirc#1(#2)[#3]{{\Lengthunit=#2\Lengthunit\up@down@{#3}%
\if@up\smov(0,.5){\arc[-][-]\arc[+][-]}\fi%
\if@dwn\smov(0,-.5){\arc[-][+]\arc[+][+]}\fi%
\def\lft{\smov(0,.5){\arc[-][-]}\smov(0,-.5){\arc[-][+]}}%
\def\rght{\smov(0,.5){\arc[+][-]}\smov(0,-.5){\arc[+][+]}}%
\if#3l\lft\fi\if#3L\lft\fi\if#3r\rght\fi\if#3R\rght\fi%
\n@ast@{#1}%
\ifnum\n@ast>0\if@up\shade[+]\fi\if@dwn\shade[-]\fi\fi%
\ifnum\n@ast>1\if@up\dshade[+]\fi\if@dwn\dshade[-]\fi\fi}}
\def\halfdashcirc(#1)[#2]{{\Lengthunit=#1\Lengthunit\up@down@{#2}%
\if@up\smov(0,.5){\dasharc[-][-]\dasharc[+][-]}\fi%
\if@dwn\smov(0,-.5){\dasharc[-][+]\dasharc[+][+]}\fi%
\def\lft{\smov(0,.5){\dasharc[-][-]}\smov(0,-.5){\dasharc[-][+]}}%
\def\rght{\smov(0,.5){\dasharc[+][-]}\smov(0,-.5){\dasharc[+][+]}}%
\if#2l\lft\fi\if#2L\lft\fi\if#2r\rght\fi\if#2R\rght\fi}}
\def\halfwavecirc(#1)[#2]{{\Lengthunit=#1\Lengthunit\up@down@{#2}%
\if@up\smov(0,.5){\wavearc[-][-]\wavearc[+][-]}\fi%
\if@dwn\smov(0,-.5){\wavearc[-][+]\wavearc[+][+]}\fi%
\def\lft{\smov(0,.5){\wavearc[-][-]}\smov(0,-.5){\wavearc[-][+]}}%
\def\rght{\smov(0,.5){\wavearc[+][-]}\smov(0,-.5){\wavearc[+][+]}}%
\if#2l\lft\fi\if#2L\lft\fi\if#2r\rght\fi\if#2R\rght\fi}}
\def\Circle#1(#2){\halfcirc#1(#2)[u]\halfcirc#1(#2)[d]\n@ast@{#1}%
\ifnum\n@ast>0\clap{%
\dimen0=\xscale\Lengthunit\vrule width#2\dimen0 height.1pt}\fi}
\def\wavecirc(#1){\halfwavecirc(#1)[u]\halfwavecirc(#1)[d]}
\def\dashcirc(#1){\halfdashcirc(#1)[u]\halfdashcirc(#1)[d]}
%
\def\xscale{1}
\def\yscale{1}
\def\Ellipse#1(#2)[#3,#4]{\def\xscale{#3}\def\yscale{#4}%
\Circle#1(#2)\def\xscale{1}\def\yscale{1}}
\def\dashEllipse(#1)[#2,#3]{\def\xscale{#2}\def\yscale{#3}%
\dashcirc(#1)\def\xscale{1}\def\yscale{1}}
\def\waveEllipse(#1)[#2,#3]{\def\xscale{#2}\def\yscale{#3}%
\wavecirc(#1)\def\xscale{1}\def\yscale{1}}
\def\halfEllipse#1(#2)[#3][#4,#5]{\def\xscale{#4}\def\yscale{#5}%
\halfcirc#1(#2)[#3]\def\xscale{1}\def\yscale{1}}
\def\halfdashEllipse(#1)[#2][#3,#4]{\def\xscale{#3}\def\yscale{#4}%
\halfdashcirc(#1)[#2]\def\xscale{1}\def\yscale{1}}
\def\halfwaveEllipse(#1)[#2][#3,#4]{\def\xscale{#3}\def\yscale{#4}%
\halfwavecirc(#1)[#2]\def\xscale{1}\def\yscale{1}}
\catcode`\@=\the\CatcodeOfAtSign

\title{Two-dimensional Lorentz-violating Chern-Simons-like action}

\author{E. Passos, A. Yu. Petrov}

\affiliation{Departamento de F\'{\i}sica, Universidade Federal da Para\'{\i}ba\\
 Caixa Postal 5008, 58051-970, Jo\~ao Pessoa, Para\'{\i}ba, Brazil}
\email{passos, petrov@fisica.ufpb.br}

\begin{abstract}
We demonstrate generation of the two-dimensional Chern-Simons-like Lorentz-breaking action via an appropriate Lorentz-breaking coupling of scalar and spinor fields at zero as well as at the finite temperature and via the noncommutative fields method and study the dispersion relations corresponding to this action.
\end{abstract}

\maketitle
\newpage
\section{Introduction}
During recent years, different aspects of possibility of the Lorentz symmetry breaking and its possible implications call high scientific interest. One of the most important directions of these investigations is the study of generation and physical impacts of new, Lorentz-breaking terms in the Lagrangian, which firstly was carried out in \cite{JK} where the new, gauge-invariant but Lorentz-breaking term was introduced into the action of electrodynamics. Further, a lot of new effects implied by the Lorentz symmetry breaking were discovered, such as ambiguity of finite quantum corrections \cite{J1,PV}, birefringence of the electromagnetic waves in vacuum, generation of this term and of its non-Abelian generalization via different methods (some papers describing these results are given in \cite{list,list1}). The Lorentz-breaking terms were also obtained and studied in the gravity \cite{JaPi,ourgra}.

At the same time, the two-dimensional field theories, due to their relatively simple structure, being motivated, in particular, by string theory, black holes theory and possibility of finding the exact solutions in certain cases, represent themselves as a convenient laboratory for studying different physical phenomena (studies of different aspects of two-dimensional field theories are presented f.e. in \cite{2d}). Thus, the very interesting problem is the possibility of generating the Lorentz-breaking terms in lower dimensions. An example of the possible Lorentz-breaking term with no higher derivatives in three-dimensional space-time is given by the mixed scalar-vector term studied earlier in the context of Julia-Toulouse mechanism \cite{JT}. Moreover, recently a two-dimensional Lorentz-breaking term was suggested and studied in the context of the defect structures \cite{BBM}. Also, it is naturally to expect, that this term naturally arises in the process of dimensional reduction of the above-mentioned three-dimensional Lorentz-breaking term and, consequently, of the four-dimensional Lorentz-breaking term arising in electrodynamics \cite{JK} which was earlier shown \cite{Ferr} to give the three-dimensional Lorentz-breaking term from \cite{JT}. In this paper, we will discuss different issues related to this term, that is, its generating via coupling to some spinor matter (both at zero and finite temperature) and the dispersion relations in theories involving such a term.

The paper is organized as follows. In the Section 2, we introduce the two-dimensional Chern-Simons-like Lorentz-breaking term
and study the dispersion relations in the theory with such a term. The Section 3 is devoted to obtaining of this term via perturbative  approach, and in the Section 4 these calculations are generalized to the finite-temperature case. The Section 5 is devoted to generation of this term via the noncommutative fields method. In the Summary, the results are discussed.

\section{Scalar field model with the Chern-Simons-like term}
In this section we formulate the two-dimensional scalar field model with the Chern-Simons-like term. The Lagrangian for this theory has the form \cite{BBM}:
\be\label{L1}
{\cal L}=-\frac{1}{2}\partial_{\mu}\phi\partial^{\mu}\phi-\frac{1}{2}\partial_{\mu}\chi\partial^{\mu}\chi+
{\rm k}_{\mu}\epsilon^{\mu\nu}\phi\partial_{\nu}\chi-V(\phi,\chi).
\ee
So, we suggest that the Lorentz violation is concentrated in the "mixed" quadratic term, similarly to the three-dimensional term \cite{JT}, whereas the potential is Lorentz invariant. Here the ${\rm k}_{\mu}$ is a vector implementing the Lorentz symmetry breaking.

We can obtain the propagators for this theory. Supposing the theory to be massless (the generation for the massive case is straightforward), we find that the action, after Fourier transform, is characterized by the matrix
\bea
\Delta=\left(\begin{array}{cc}
-p^2 & i{\rm k}_{\mu}\epsilon^{\mu\nu}p_{\nu}\\
-i{\rm k}_{\mu}\epsilon^{\mu\nu}p_{\nu} &-p^2
\end{array}
\right)=-p^2{\mathbf 1}-{\rm k}_{\mu}\epsilon^{\mu\nu}p_{\nu}\sigma_2,
\eea
where $\sigma_2=\left(\begin{array}{cc}
0 & -i\\
i & 0
\end{array}
\right)$ is a corresponding Pauli matrix. 

The matrix propagator is given by the matrix inverse to $\Delta$, that is
\bea
\left(\begin{array}{cc}
<\phi\phi> & <\phi\chi>\\
<\chi\phi> &<\chi\chi>
\end{array}
\right)=-\frac{p^2}{p^4+p^2{\rm k}^2-(p\cdot {\rm k})^2}\left(1-\frac{{\rm k}_{\mu}\epsilon^{\mu\nu}p_{\nu}}{p^2}\sigma_2\right).
\eea
The dispersion relations can be found from the poles of the denominator of this expression. For the signature $diag(+-)$ they look like (here $E=p_0,\vec{p}=p_1$):
\bea
(E^2-\vec{p}^2)^2+(E^2-\vec{p}^2){\rm k}^2-(E{\rm k}_0-\vec{p}{\rm k}_1)^2=0.
\eea
There are three characteristic cases:

\noindent (i) Time-like ${\rm k}^{\mu}$, ${\rm k}_0\neq 0$, ${\rm k}_1=0$ gives dispersion relation $E^2-\vec{p}^2=\pm|\vec{p}||{\rm k}_0|$ for which we have birefringence. Indeed, the group velocities of the possible waves are $v_{gr}=\frac{\vec{p}\pm\frac{1}{2}|{\rm k}_0|}{\sqrt{\vec{p}^2\pm |\vec{p}||{\rm k}_0|}}$. For both signs these group velocities are greater than the speed of light, which allows to treat this case as a nonphysical one. \\
(ii) Space-like ${\rm k}^{\mu}$, ${\rm k}_0=0$, ${\rm k}_1\neq 0$ gives dispersion relation $E^2-\vec{p}^2=\pm|E||{\rm k}_1|$, so, we have birefringence due to two different phase velocities, with the group velocities $v_{gr}=\pm\frac{\vec{p}}{\sqrt{\vec{p}^2+\frac{{\rm k}^2_1}{4}}}$, and with no superluminal group velocities.\\
(iii) Light-like ${\rm k}^{\mu}$, ${\rm k}_0={\rm k}_1=k$ gives either common dispersion relation $E=\pm\vec{p}$ or the deformed dispersion relation $E-\vec{p}=\pm k$, for which the phase velocity is variable whereas the group velocity coincides with the speed of light.

\section{Generation of the Chern-Simons-like term via radiative corrections}
Let us consider the model of fermions interacting with scalar fields $\phi(x)$ and $\chi(x)$, with the Lorentz symmetry violation is implemented via a constant vector $b^{\mu}$. The Lagrangian density of the model is as follows:

\be
{\cal L}=-\frac{1}{2}\partial_{\mu}\phi\partial^{\mu}\phi-\frac{1}{2}\partial_{\mu}\chi\partial^{\mu}\chi+\bar{\psi}( i \pls - m )\psi - g\bar{\psi} \bs \phi\psi - g\,m\bar{\psi}\gamma_{5}\chi \psi.
\ee

We are planning to find the one-loop Chern-Simons-like effective action of the scalar fields. It is natural to suggest that in the momentum space it can be represented as
\bea
\Gamma=\int\frac{d^2q}{(2\pi)^2}\phi(-q)\Pi(q)\chi(q),
\eea
with the $\Pi(q)$ is the self-energy tensor.

Applying the following Feynman rules:

\vspace*{2mm}

\hspace{2cm}
\Lengthunit=1.2cm
\GRAPH(hsize=3){\Linewidth{.6pt}\lin(1,0)\ind(15,0){\;\;\;\;\;\;\;\;\;\;\;\;=\frac{i(\ps+m)}{p^{2}-m^{2}}}
}
\hspace{.5cm}
\Lengthunit=1.2cm
\GRAPH(hsize=3){\ind(5,0){\bullet}\lin(1,0)\mov(.5,0){\dashlin(0,.7)}\ind(22,0){=-ig\,m\gamma_5}
}
\hspace{.5cm}
\Lengthunit=1.2cm
\GRAPH(hsize=3){\ind(5,0){\times}\lin(1,0)\mov(.5,0){\wavelin(0,.7)}\ind(18,0){\;\;\;\;\;\;\;\;\;=-ig\bs,}
}

\vspace*{2mm}

\noindent we arrive at the following diagrams that contribute to the two-point "mixed" function of the scalar fields (it is easy to see that the Lorentz-breaking contributions to the two-point function of the $\phi$ field only or of $\chi$ field only represent themselves as total derivatives):

\vspace*{3mm}

\hspace{4.0cm}
\Lengthunit=1.2cm
\GRAPH(hsize=2){\wavelin(.5,0)\mov(1,0){\Circle(1)}\mov(1.5,0){\dashlin(.5,0)}\ind(15,0){\bullet}\ind(5,0){\times}
\ind(10,-10){(a)}
}
\hspace{.5cm}
\Lengthunit=1.2cm
\GRAPH(hsize=2){\dashlin(.5,0)\mov(1,0){\Circle(1)}\mov(1.5,0){\wavelin(.5,0)}\ind(15,0){\times}\ind(5,0){\bullet}
\ind(10,-10){(b)}
}
\vspace*{3mm}

\noindent Here the simple line is for the propagator of the $\psi$ field, the wavy line -- for the external $\phi$ field, and the dashed line -- for the external $\chi$ field.
These diagrams produce two contributions to self-energy tensor:

\bea\label{t1}
\Pi_{a}(q)=-ig^{2}\int\frac{d^{2}p}{(2\pi)^{2}}{\rm tr}\bigl[\bs\,S(p)\gamma_{5}\,S(p+q)]
\eea
and

\bea\label{t2}
\Pi_{b}(q)=-ig^{2}\int\frac{d^{2}p}{(2\pi)^{2}}{\rm tr}\bigl[\gamma_{5}\,S(p)\bs\,S(p+q)],
\eea
where $S(p)$ is the usual Dirac propagator for fermions. Summing up the self-energy tensors in (\ref{t1}) and (\ref{t2}), we can write down the following expression for the self-energy tensor $\Pi(q)$ characterizing the new, Lorentz-breaking contribution to the quadratic effective action of the theory: 
\bea\label{i1}
\Pi(q)=\Pi_a(q)+\Pi_b(q)=-2 m^{2}g^{2}\int^{1}_{0}dx (1-x)\int\frac{d^{2}p}{(2\pi)^{2}}\frac{1}{(p^{2}-M^{2})^{2}}\,b_{\mu}\epsilon^{\mu\nu}q_{\nu},
\eea
where we have used Feynman parameter $x$ to form the denominator in (\ref{i1}), with $M^{2}=m^{2}-q^{2}(1-x)x$. 
To simplify the numerator in (\ref{i1}), we use the fact that the trace of product of $\gamma_{5}$ with an even number of $\gamma$ matrices can be reduced: ${\rm tr}[\gamma^{\mu}\gamma^{\nu}\gamma_{5}]=2i\epsilon^{\mu\nu}$ and ${\rm tr}[\gamma_{5}]=0$. Also, by symmetry reasons we can omit all terms which are odd in the internal momentum $p$. The integration over the internal momentum in the self-energy tensor (\ref{i1}) gives the result:
\bea\label{i2}
\Pi(q)&=&-\frac{i m^{2}g^{2}}{2\pi}\int^{1}_{0}dx (1-x)\frac{1}{m^{2}-q^{2}(1-x)x}\,b_{\mu}\epsilon^{\mu\nu}q_{\nu}\nonumber\\&\simeq&-\frac{i g^{2}}{4\pi}\,b_{\mu}\epsilon^{\mu\nu}q_{\nu},
\eea
where we have neglected higher orders in the external momentum $q_{\nu}$.

The corresponding Chern-Simons-like effective action of the scalar fields after the Fourier transform looks like:
\bea
\Gamma=\int d^{2}x\,{\rm k}_{\mu}\epsilon^{\mu\nu}\phi\partial_{\nu}\chi,
\eea
with the following relation between the constant vectors ${\rm k}_{\mu}$ and $b_{\mu}$ takes place:
\bea
{\rm k}_{\mu}=\frac{g^{2}}{4\pi}\,b_{\mu}.
\eea
One can observe that this factor is similar to the result obtained in \cite{list1,ourd4}, up to the factor $\frac{1}{\pi}$. We note that this result is not ambiguous due to its explicit finiteness (see discussion of ambiguities in the Lorentz-breaking theories in \cite{ourd4,J1,PV}).

\section{Finite temperature case} 

Let us implement the finite temperature in this theory. To do it, first we carry out the Wick rotation in (\ref{i1}), and use a Matsubara formalism for fermions, which consists in taking $p_{0}\equiv \omega_n=(n+1/2)\frac{2\pi}{\beta}$ and replacement of the integration over zeroth component of the momentum by a discrete sum by the rule $\frac{1}{2\pi}\int dp_{0}\rightarrow\frac{1}{\beta}\sum_{n}$ \cite{Dolan}. Thus, the self-energy tensor $\Pi(q)$ (\ref{i1}), which determines the correction to the quadratic action and represents itself as the key object of the method, takes the form
\bea\label{i1t}
\Pi(q)=-\frac{im^2g^2}{\beta}\int\frac{dp_1}{2\pi}\sum\limits_{n=-\infty}^{+\infty}\frac{1}{[\frac{4\pi^2}{\beta^2}(n+\frac{1}{2})^2 +p^{2}_1+m^{2}]^{2}}\,b_{\mu}\epsilon^{\mu\nu}q_{\nu},
\eea
which after integration yields
\bea
\label{sum}
\Pi(q)=-\frac{ig^2a^2}{8\pi}b_{\mu}\epsilon^{\mu\nu}q_{\nu}\sum\limits_{n=-\infty}^{+\infty}\frac{1}{[(n+b)^2 +a^{2}]^{\lambda}},
\eea
where $a=\frac{m\beta}{2\pi}$, $b=1/2$ and $\lambda=3/2$.

\begin{figure}[ht]
\centerline{\includegraphics[{angle=90,height=7.0cm,angle=180,width=8.0cm}]
{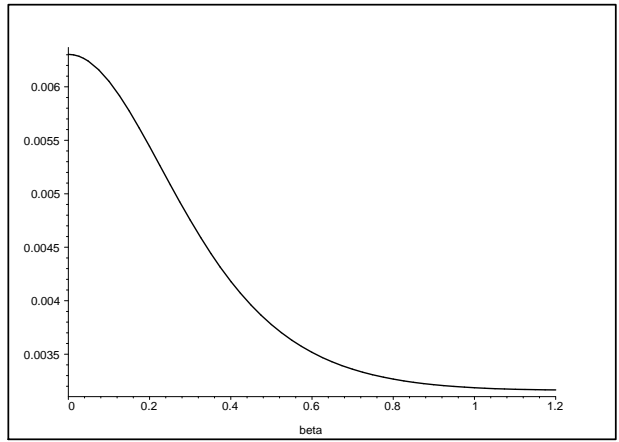}} \caption{The function $F(a)$ is different from
zero everywhere. At zero temperature ($\beta\to\infty$), the
function tends to a nonzero value.}\label{fig2}
\end{figure}

The sum (\ref{sum}) is evidently finite. To calculate it, we proceed in the following way. 
First of all, we apply an explicit expression for the sum over the Matsubara frequencies \cite{FO}:
\bea\label{fo}
\sum_n \frac{1}{[(n+b)^2 + a^2]^{\lambda}}= \frac{\sqrt{\pi}\Gamma(\lambda
- 1/2)}{\Gamma(\lambda)(a^2)^{\lambda - 1/2}}
+
4\sin(\pi\lambda)\int_{|a|}^\infty \frac{dz}{(z^2 - a^2)^{\lambda}}
Re\left(\frac{1}{\exp 2\pi(z + ib) -1}\right).
\eea 
However, this expression is valid for $1/2 <\lambda <1$, therefore it cannot be straightforwardly applied for $\lambda=3/2$ since the integral in the right-hand side of this expression diverges. So, we must proceed similarly to \cite{ourd4}, that is, we carry out the analytic continuation of the second term of the right-hand side of this identity, see \cite{grig,ourd4}:
\bea
&&\int_{|a|}^\infty \frac{dz}{(z^2 - a^2)^{\lambda}}
Re\left(\frac{1}{\exp 2\pi(z + ib) -1}\right)= 
\frac{1}{2a^2}\frac{3-2\lambda}{1-\lambda}
\int_{|a|}^\infty \frac{dz}{(z^2 - a^2)^{\lambda-1}}\times\\&\times&
Re\left(\frac{1}{\exp 2\pi(z + ib) -1}\right)-\nonumber\\  &-&
\frac{1}{4a^2}\frac{1}{(2-\lambda)(1-\lambda)} \int_{|a|}^\infty
\frac{dz}{(z^2 - a^2)^{\lambda-2}}
\frac{d^2}{dz^2}Re\left(\frac{1}{\exp 2\pi(z + ib) -1}\right).\nonumber 
\eea
This expression can be used for $\lambda=\frac{3}{2}$ and gives no singularities in this case. Then, putting all together, we get
\bea
\Pi_{a}(q)&=&-\frac{ig^2}{4\pi}b_{\mu}\epsilon^{\mu\nu}q_{\nu}\Big[1+
2\pi^2F(a)\Big],
\eea
and, consequently,
\bea
{\rm k}_{\mu}=\frac{g^2}{4\pi}b_{\mu}\Big[1+
2\pi^2F(a)\Big]
\eea
where the function
\bea
F(a)=\int_{|a|}^\infty dz(z^2 - a^2)^{1/2} \frac{\tanh (\pi z)}{\cosh^2(\pi z)}
\eea
has the following asymptotics: $F(a\to\infty)\to 0$ ($T\to 0$), $F(a\to 0)\to \frac{1}{2\pi^2}$ ($T\to \infty$), see Fig.\ref{fig2}. 

Thus we see that the value of ${\rm k}_{\mu}$ at the high temperature is its doubled value at the low temperature. The similar temperature dependence of the Lorentz-breaking parameter has been observed earlier for the four-dimensional electrodynamics \cite{ourd4}.

\section{Noncommutative fields method}

The noncommutative fields method is known to be an efficient mechanism of introducing the Lorentz-breaking terms into the action \cite{ourgra,Gamb}. The essence of this method consists in deformation of the canonical commutation relations implying in arising of new additive, in general case Lorentz-breaking terms in the Lagrangian. Earlier \cite{ourgra,Gamb} this method was mostly applied to the gauge theories where it was based on deformation of the constraints arisen by requirements of the gauge symmetry. However, as we will see here, despite there is no constraints in the scalar field theory, 
in this case deformation of the canonical commutation relations will also modify the Lagrangian of the theory.

Let us for convenience consider the theory of the complex scalar field:
\bea
{\cal L}=-\pa_{\mu}\sigma^*\pa^{\mu}\sigma-V(\sigma,\sigma^*).
\eea
This theory is equivalent to the theory (\ref{L1}), under change $\sigma=\phi+i\chi$, $\sigma^*=\phi-i\chi$.
The canonical momenta are $\pi=-\dot{\sigma}^*$, $\pi^*=-\dot{\sigma}$. The corresponding Hamiltonian density is
\bea
H=-\pi\pi^*-\pa_1\sigma\pa_1\sigma^*+V(\sigma,\sigma^*)
\eea
To quantize the theory in general case, we must convert fields and momenta into the operators and employ the canonical commutation relations: $[\sigma(\vec{x}),\pi(\vec{y})]=i\delta(\vec{x}-\vec{y})$, $[\sigma^*(\vec{x}),\pi^*(\vec{y})]=i\delta(\vec{x}-\vec{y})$, with all other commutators of the canonical variables are equal to zero, which evidently reproduces the known Hamilton equations. 

Now, let us implement the noncommutative fields method. This is done by modification of only one commutation relation:
\bea
[\pi(\vec{x}),\pi^*(\vec{y})]=\theta\delta(\vec{x}-\vec{y}),
\eea   
with $\theta$ is a some constant. In this case, the $\sigma$ and $\sigma^*$ continue to be the canonical variables whereas $\pi$ and $\pi^*$ do not more. So, we are to introduce the new canonical variables $\Pi=\pi+i\frac{\theta}{2}\sigma^*$, $\Pi^*=\pi^*-i\frac{\theta}{2}\sigma$. The only modification of the Lagrangian, which now has the form
\bea
{\cal L}^{\prime}=\Pi\dot{\sigma}+\Pi^*\dot{\sigma}^*-H,
\eea
originates from the modification of the canonical variables (unlike of \cite{Gamb}, there is no constraints in this theory). Finally, we arrive at
\bea
{\cal L}^{\prime}={\cal L}-\frac{i}{2}\theta(\sigma\dot{\sigma}^*-\dot{\sigma}\sigma^*),
\eea
which in terms of the fields $\phi,\chi$ can be rewritten as
\bea
{\cal L}^{\prime}={\cal L}+\theta(\chi\dot{\phi}-\phi\dot{\chi}),
\eea
which exactly reproduces the Lagrangian (\ref{L1}), with the Lorentz-breaking vector ${\rm k}^{\mu}$ be ${\rm k}^{\mu}=(0,\theta)$. So, we succeeded to generate the Lorentz-breaking term in the action, with the Lorentz-breaking vector turns out to be space-like, as it follows from the analysis of the dispersion relations.

\section{Summary}

We succeeded to construct the two-dimensional Lorentz-breaking term. This term turns out to modify the dispersion relations, with the physically consistent cases are those ones with the space-like or light-like Lorentz-breaking vector. We also have generated this term via an appropriate, Lorentz-breaking coupling of the scalar fields with the spinor ones, both in the zero temperature case and finite temperature case, with this term turns out to grow with the temperature.  Also, we succeeded to generate the Lorentz-breaking term using the noncommutative fields method, with the Lorentz-breaking vector turns out to be space-like, that is, consistent with the dispersion relations. We expect that this term can be applied for further studies of the topological defects, similarly to the studies in \cite{BBM}.

{\bf Acknowledgements.} Authors are grateful to Prof. D. Bazeia for calling their attention to this model. This work was partially supported by Conselho Nacional de Desenvolvimento Cient\'{\i}fico e Tecnol\'{o}gico (CNPq). The work by A. Yu. P. has been supported by CNPq-FAPESQ DCR program, CNPq project No. 350400/2005-9.

\end{document}